# Destruction of superconductivity through phase fluctuations in ultrathin *a*-MoGe films

Soumyajit Mandal[a], Surajit Dutta[a], Somak Basistha[a], Indranil Roy[a], John Jesudasan[a], Vivas Bagwe[a], Lara Benfatto[b] and Arumugam Thamizhavel[a], Pratap Raychaudhuri[a] [1]

[a] *Tata Institute of Fundamental Research, Homi Bhabha Road, Colaba, Mumbai 400005, India.*

[b] *Department of Physics, Sapienza University of Rome, Piazzale Aldo Moro, 5, 00185 Rome, Italy.*

We investigate the evolution of superconductivity with decreasing film thickness in ultrathin amorphous MoGe (*a*-MoGe) films using a combination of sub-Kelvin scanning tunneling spectroscopy, magnetic penetration depth measurements and magneto-transport measurements. We observe that superconductivity is strongly affected by quantum and classical phase fluctuations for thickness below 5 nm. The superfluid density is strongly suppressed by quantum phase fluctuations at low temperatures and evolves towards a linear-$T$ dependence at higher temperatures. This is associated with a rapid decrease in the superconducting transition temperature, $T_c$, and the emergence of a pronounced pseudogap above $T_c$. These observations suggest that at strong disorder the destruction of superconductivity follows a Bosonic route where the global superconducting state is destroyed by phase fluctuations even though the pairing amplitude remains finite.

[1] E-mail: pratap@tifr.res.in

In the late fifties, Anderson[1] predicted that in an s-wave superconductor the attractive pairing interaction forming Cooper pairs would remain largely unaffected by the presence of non-magnetic impurities. This has been loosely interpreted to imply that the superconducting transition temperature, $T_c$ will also not be strongly sensitive to disorder scattering. However, later experiments showed that this is valid only in the limit of weak disorder: In the presence of strong disorder, $T_c$ gets gradually suppressed[2,3] and eventually the material is driven into a non-superconducting state. The mechanism driving the transition from a superconductor to an insulator or a metal has been a subject of considerable debate. In principle, the suppression of $T_c$ with increase in disorder can happen from two origins. The first mechanism is through the loss of effective screening with increase in disorder that weakens the attractive pairing interaction, and thus suppresses the mean field transition temperature[4,5]. The second mechanism results from the decrease in superfluid density, $n_s$, induced by disorder scattering, which renders the superconductor susceptible to phase fluctuations[6,7]. When $n_s$ is small, the phase coherent superconducting state can get destroyed due to strong phase fluctuations even when the pairing amplitude remains finite[8,9,10].

The superconductor to non-superconductor transition driven by these two mechanisms are often classified as the Fermionic and Bosonic routes respectively[11]. In the Fermionic route, the pairing attraction drops to zero at a critical disorder where superconductivity is destroyed. This non-superconducting state is either a bad metal or an Anderson insulator. In the Bosonic mechanism the pairing interaction remains finite and therefore signatures of Cooper pairing continue to survive even in the non-superconducting state. Experimentally, this manifests as a persistence of the superconducting gap in the electronic excitation spectrum, known as the pseudogap, even after the global superconducting state is destroyed[12,13,14,15]. However, recent studies indicate that this classification may be oversimplified: The same system can follow the Fermionic route at moderate disorder and cross-over to a Bosonic scenario at stronger disorder[16,17]. It is therefore interesting to investigate whether a system can follow the Fermionic route all the way to the disorder level where the superconducting ground state is completely destroyed.

The amorphous superconductor *a*-MoGe is widely believed to follow the Fermionic route till the destruction of superconductivity[18,19,20]. The very short electronic free path which is much smaller than the coherence length, ξ, puts this system in the extreme dirty limit. Furthermore, effect of disorder in *a*-MoGe films gets accentuated with decrease in thickness making it possible to investigate the suppression of superconductivity by using thickness as the tuning parameter. Detailed transport measurements relating the sheet resistance with $T_c$ appear to be consistent with the Fermionic theory proposed by Finkelstein[4]. Nevertheless, the presence of quantum phase fluctuations at very low thickness has also been recognised[21]. Thus, it is important to carry out detailed measurements with more sophisticated probes which can resolve the presence of a pseudogap to confirm whether the system follows the Fermionic route down to very low thickness.

In this letter, we investigate the evolution of superconductivity in ultrathin *a*-MoGe films, using a combination of low temperature scanning tunnelling spectroscopy (STS), penetration depth ($\lambda$) and magnetotransport measurements. We observe that for the film with thickness down to 5 nm, the decrease in $T_c$ appears to be consistent with the expectation from Fermionic theories. However, below this thickness, $T_c$ decreases rapidly with a rapid increase in $\Delta/k_BT_c$ ratio. The films show strong signature of quantum and classical phase fluctuations and a pronounced pseudogap from STS measurements, suggesting that the eventual destruction of superconductivity follows the Bosonic route.

The *a*-MoGe thin films used in this study were grown on oxidised silicon substrates using pulsed laser deposition starting from a $Mo_{70}Ge_{30}$ arc-melted target. Details of sample preparation have earlier been reported in refs. 22, 23. Film thickness ($t$) was varied between 20 nm to 1.8 nm. For $t \geq 10$ nm the thickness of the film was directly measured using stylus profilometer whereas for thinner samples it was estimated from the number of laser pulses using two films with $t \geq 10$ nm grown before and after the actual run for calibration. Magnetotransport and penetration depth measurements were performed in $^3$He cryostats operating down to 300 mK. STS measurements were performed using a home-built scanning tunnelling microscope[24] (STM) operating down to 450 mK and in magnetic fields up to 90 kOe. The tunnelling conductance ( $G(V) = \left.\frac{dI}{dV}\right|_V$ $vs\ V$ ) was measured using standard modulation technique using a Pt-Ir tip. To maintain a pristine surface, the sample used for STS measurement was transferred in an ultrahigh vacuum suitcase after deposition and transferred in the STM without exposure to air. The penetration depth was measured using a low frequency (30 kHz) two-coil mutual inductance technique[25,26,27] that allows the determination of the absolute value of the penetration depth in thin films.

Fig. 1(a) shows the sheet resistance, $R_{\square}$, as a function of temperature for samples with different thicknesses. We define $T_c$ as the temperature where the resistance becomes < 0.05% of the normal state value. Fig. 1(b) shows the variation of the normal state sheet resistance (taken as the sheet resistance at 9 K, $R_{\square}^{9K}$ ) and $T_c$ as a function of film thickness. With decreasing thickness $T_c$ decreases whereas $R_{\square}$ increases, but remains well below the quantum resistance, $\frac{h}{4e^2}$ = 6.45 kΩ (where $e$ is the electron charge and $h$ is the Planck's constant). For $t \geq 8$ nm $R_{\square}^{9K}$ varies linearly with $1/t$ showing that the increase in the sheet resistance is primarily a geometric effect ( *inset* of Fig. 1(b) ). Below this thickness $R_{\square}^{9K}$ shows an upward trend and the corresponding resistivity ( $\rho_N$) shows an increase from approximately 1.5 μΩ-m to 2.6 μΩ-m.

We first concentrate on the STS measurements. *G(V)-V* spectra recorded over a 32 ×32 grid over 200 nm × 200 nm area at each temperature. Fig. 2(a) and 2(b) show the average spectra at different temperatures for the 20 nm and 2 nm thick samples. At low temperature the spectra for all samples have the characteristic features of a superconductor: a depression in *G(V)* at low bias corresponding to the superconducting energy gap, Δ, and the presence of coherence peaks at the gap edge. In addition, for the 2 nm thick sample, we observe a broad V-shaped, nearly temperature independent background. This feature, also observed in other disordered superconductors[28,29] is attributed to the Altshuler-Aronov type electron-electron interactions in disordered metals. To extract the superconducting contribution alone, we calculated the normalised spectra, $G_N(V)$ vs. $V$, by dividing it with the spectra obtained at high temperature where the low bias feature associated with superconducting pairing disappears. The left panels of Fig. 2(c)-(e) show the temperature dependence of $G_N(V)$ vs. $V$ spectra for three representative films in the form of intensity plots along with the temperature variation of $R_{\square}$. For the samples with $t \sim 11$ nm the superconducting energy gap closes very close to $T_c$ consistent with the expectation from Bardeen-Cooper-Schrieffer theory[30]. For the sample with $t \sim 4.5$ nm we see a small hint of a pseudogap, where soft gap in the tunneling spectra that extends approximately 0.5 K above $T_c$. The pseudogap regime gets extended for the sample with $t \sim 2$ nm to about double of $T_c$. We define the pseudogap temperature, $T^*$, as the temperature where $G_N(0) \approx 0.95$. At the same time, the zero bias conductance ( $G_N(0)$ ) maps obtained at 450 mK ( right panels of Fig. 2(c)-(e) ) reveal that the superconducting state becomes progressively inhomogeneous as we go to lower thickness. Plotting $T_c$ and $T^*$ as a function of film thickness (Fig. 2(f)), we observe that a large region of pseudogap emerges below a thickness of 4 nm. At the same time, extracting $\Delta$ at 450 mK using the BCS + $\Gamma$ model[31,32], we observe that $\Delta/(k_B T_c)$ increases rapidly below $t \sim 5$ nm reaching a value of 5 at 2 nm. The emergence of pseudogap between $T_c$ and $T^*$ as well as the anomalously large value of $\Delta/(k_B T_c)$ both signal the breakdown of the BCS scenario in very thin $a$-MoGe films.

We now look at the penetration depth data. Fig. 3(a) shows the temperature variation of $\lambda^{-2}$ for different films. At low temperatures $\lambda^{-2}$ saturates towards a constant value for all samples. We first analyze the thickness variation of $\lambda^{-2}(T \rightarrow 0)$. With decrease in thickness $\lambda^{-2}(T \rightarrow 0)$ progressively decreases by more than an order of magnitude. Within BCS theory, with increase in disorder, $n_s$ ($\equiv \frac{m^*}{\mu_0 e^2 \lambda^2}$, where $m^*$ is the effective mass and $\mu_0$ is the vacuum permeability) gets suppressed from the electronic carrier density, $n$, due to increase in electron scattering. In the dirty limit, this is captured by the relation[33], $\lambda_{BCS}^{-2}(0) = \frac{\pi \mu_0 \Delta(T \rightarrow 0)}{\hbar \rho_N}$, where $\mu_0$ is the vacuum permeability, $\hbar = \frac{h}{2\pi}$ is the reduced Planck's constant and $\rho_N$ is the resistivity in the normal state ( at 10 K ). Plotting $\frac{\lambda^{-2}(T \rightarrow 0)}{\lambda_{BCS}^{-2}(0)}$ ( Fig. 3(b) ), we observe that the measured $\lambda^{-2}$ falls significantly below the disorder suppressed BCS value below $t \sim 5$ nm, reaching a value $\sim 0.55$ for 2.2 nm thick sample. Coming to the temperature dependence, we note that the curves for samples with $t > 5$ nm can be fitted with the approximate dirty limit BCS expression, $\frac{\lambda^{-2}(T)}{\lambda^{-2}(0)} = \frac{\Delta(T)}{\Delta(0)} \tanh\left[\frac{\Delta(T)}{2k_B T}\right]$ ( where $k_B$ is the Boltzmann constant ), where $\Delta(0)$ is constrained within 10% of the value obtained from tunneling measurements at 450 mK and $\Delta(T)/\Delta(0)$ is assumed to have the BCS temperature dependence[31] for a weak coupling s-wave superconductor. However, for the sample with $t$ = 2.2 nm the qualitative nature of the temperature variation is different ( Fig. 3(c) ): $\lambda^{-2}$ shows a slow variation at low temperature and crosses over to a linear variation before decreasing rapidly close to $T_c$.

Since the suppression of $\lambda^{-2}(0)$ and the linear-$T$ variation are consistent with the expectations from quantum and classical longitudinal phase fluctuations[34,35], we now attempt a comparison of these features with theory[36,37,38]. The resilience of a superconductor against phase fluctuations is given by the superfluid stiffness, which for a 2-dimentional superconductor ( $t \ll \xi$ ) is given by, $J_s = \frac{\hbar^2 n_s t}{4m^*}$. For the 2.2 nm thick sample, we estimate $\xi \sim 8\ nm$ from $H_{c2}$, such that it is in the 2D limit. A rough estimate of the suppression of $n_s$ due to quantum phase fluctuations can be obtained by considering two energy scales[39]: The Coulomb energy, $E_C = \frac{e^2}{2\epsilon_0 \epsilon_B \xi}$ and $J_s(T \to 0)$, where $\epsilon_0$ is the vacuum permeability and $\epsilon_B$ is the background dielectric constant. Using the carrier density measured from Hall effect measurements $n = 4.63 * 10^{29}\ m^{-3}$ and the plasma frequency[40] $\Omega_p = 1.625 * 10^{16}\ Hz$, we estimate $\epsilon_B = \frac{e^2 n}{\epsilon_0 m \Omega_p^2} \sim 5.6$. Here, the suppression of the superfluid density due to quantum phase fluctuations is given by[31], $\frac{n_s}{n_s^0} = Exp\left[-\frac{\langle(\Delta\theta)^2\rangle}{4}\right]$, where $\langle(\Delta\theta)^2\rangle \approx \frac{1}{5\pi}\sqrt{\frac{E_C}{J_s(0)}}$, and $n_s^0$ is the bare superfluid density in the absence of phase fluctuations. We obtain $\frac{n_s}{n_s^0} \sim 0.78$. This value would get further reduced in a disordered system if the local superfluid density is spatially inhomogeneous[41,42,43,44]. Though it is difficult to quantify this effect in our film, from the large spatial variation of $G_N(0)$ in the 2 nm thick films we believe that this effect could be substantial. This additional suppression of the superfluid density due to inhomogeneity reflects in the emergence of a finite-frequency absorption that is expected to occur at relatively low energies, i.e. below the 2Δ threshold for quasiparticle absorption in dirty superconductors[45,46]. This effect has been observed in NbN and $InO_x$ films[47,48]. Such low energy dissipative mode has a feedback on the spectrum of the quantum phase fluctuations such that the effect of quantum corrections gets reduced and the quantum to classical cross-over shifts to a lower temperature[39]. Considering these uncertainties at the moment we can only state that the observed value of $\frac{\lambda^{-2}(300\ mK)}{\lambda_{BCS}^{-2}(T\to 0)} \sim 0.55$ falls in the correct ballpark.

Finally, to reconcile the penetration depth measurements with the emergence of pseudogap, we can now compare two energy scales, the pairing energy $\Delta(0)$ and superfluid stiffness, $J_s(0)$. The superconducting $T_c$ is determined by the lower of these two energy scales[6]. When $\Delta(0) << J_s(0)$, the superconducting transition temperature is given by, $T_c \sim \frac{\Delta(0)}{Ak_B}$, where[40] $A \sim 2$ for *a*-MoGe. On the other hand when $J_s(0) << \Delta(0)$, thermal phase fluctuations play a dominant role. Here the superconducting state gets destroyed due to thermal phase fluctuations at $T_c \sim \frac{J_s(0)}{Bk_B}$ ( $B$ is a constant of the order of unity ) even if the pairing amplitude remains finite up to higher temperatures. From Fig. 3(d) we observe that for films with $t$ > 5nm, $J_s(0)$ is one order of magnitude larger than $\Delta(0)$. However, below 5 nm $J_s(0)$ decreases rapidly and around $t \sim 2$ nm both become of the same order. Thus around this thickness we expect phase fluctuation to dominate the superconducting properties. The observation of the pseudogap suggests that at this thickness the pairing amplitude remains finite even when the global phase coherent state has been destroyed by phase fluctuations. In addition since the film is in the 2D limit, the measured temperature dependence of the superfluid stiffness near $T_c$ turns out to be consistent[31] with a Berezinskii-Kosterlitz-Thouless[49,50,51,52,53] (BKT) jump smeared by disorder-induced inhomogeneity[54,55].

In summary, we have shown that the suppression of superconductivity in *a*-MoGe with decreasing film thickness has two regimes. At moderate disorder $T_c$ decreases but $\Delta/k_BT_c$ shows only a small increase consistent with Fermionic theories. At stronger disorder the system crosses over to a Bosonic regime where the pairing amplitude remains finite even after the superconducting state is destroyed by phase fluctuations. This evolution of the superconducting state is consistent with earlier observations on disordered[12,13,14] NbN,

TiN and $InO_x$ where the eventual destruction of superconductivity through the Bosonic route is well established. The observation of Bosonic scenario in $a$-MoGe, which was widely believed to follow the Fermionic route raises the important question on whether a disordered superconductor can follow the Fermionic route all the way to the destruction of superconductivity or whether all superconductors become Bosonic at sufficiently strong disorder. This question needs to be addressed in future theoretical studies.

We thank T. V. Ramakrishnan for valuable discussions. This work was financially supported by Department of Atomic Energy, Government of India (Grant No. 12-R&D-TFR-5.10-0100), the Italian MAECI and the Department of Science and Technology, Government of India under the Italy-India collaborative project (SUPERTOP-PGR04879 and INT/Italy/P-21/2016 (SP)), by the Italian MIUR project PRIN 2017 n.2017Z8TS5B, and by Regione Lazio (L.R. 13/08) under project SIMAP.

SM along with SB performed the penetration depth measurements and analyzed the data. SD and IR performed the scanning tunneling spectroscopy measurements and analyzed the data. SD and SB performed the transport measurements. JJ, VB and AT prepared the bulk target and thin films and performed preliminary characterization. LB provided theoretical inputs. PR conceived the problem, supervised the project and wrote the paper with inputs from all authors.

---

[20] A. M. Goldman, *Superconductor-Insulator Transitions*, Int. J. Mod. Phys. B **24,** 4081 (2010).

[21] A. Yazdani and A. Kapitulnik, *Superconducting-Insulating Transition in Two-Dimensional a-MoGe Thin Films*, Phys. Rev. Lett. **74**, 3037 (1995).

[22] I. Roy, S. Dutta, A. N. Roy Choudhury, S. Basistha, I. Maccari, S. Mandal, J. Jesudasan, V. Bagwe, C. Castellani, L. Benfatto, and P. Raychaudhuri, *Melting of the Vortex Lattice through Intermediate Hexatic Fluid in an a-MoGe Thin Film***,** Phys. Rev. Lett. **122**, 047001 (2019).

[23] S. Dutta, I. Roy, S. Basistha, S. Mandal, J. Jesudasan, V. Bagwe and P. Raychaudhuri, *Collective flux pinning in hexatic vortex fluid in a-MoGe thin film*, J. Phys.: Condens. Matter **32**, 075601 (2020).

[24] A. Kamlapure, G. Saraswat, S. C. Ganguli, V. Bagwe, P. Raychaudhuri, and S. P. Pai, *A 350 mK, 9 T scanning tunneling microscope for the study of superconducting thin films on insulating substrates and single crystals*, Rev. Sci. Instrum. **84**, 123905 (2013).

[25] A. Kamlapure, M. Mondal, M. Chand, A. Mishra, J. Jesudasan, V. Bagwe, L. Benfatto, V. Tripathi and P. Raychaudhuri, *Measurement of magnetic penetration depth and superconducting energy gap in very thin epitaxial NbN films*, Appl. Phys. Lett. **96**, 072509 (2010).

[26] S. J. Turneaure, E. R. Ulm, and T. R. Lemberger, *Numerical modeling of a two-coil apparatus for measuring the magnetic penetration depth in superconducting films and arrays,* J. Appl. Phys. **79**, 4221 (1996).

[27] S. J. Turneaure, A. A. Pesetski, and T. R. Lemberger, *Numerical modeling and experimental considerations for a two-coil apparatus to measure the complex conductivity of superconducting films,* J. Appl. Phys. **83**, 4334 (1998).

[28] S. P. Chockalingam, M. Chand, A. Kamlapure, J. Jesudasan, A. Mishra, V. Tripathi and P. Raychaudhuri, *Tunneling studies in a homogeneously disordered s-wave superconductor: NbN***,** Phys. Rev. B **79**, 094509 (2009).

[29] G. Lemarie, A. Kamlapure, D. Bucheli, L. Benfatto, J. Lorenzana, G. Seibold, S. C. Ganguli, P. Raychaudhuri, and C. Castellani, *Universal scaling of the order-parameter distribution in strongly disordered superconductors*, Phys. Rev. B **87**, 184509 (2013).

[30] M. Tinkham, Introduction to Superconductivity (McGrawHill, Singapore, 1996).

[31] See Supplemental Material for (i) Fitting of the tunneling spectra and temperature variation of superconducting energy gap; (ii) Two-coil mutual inductance technique for penetration depth measurements; and (iii) Derivation of formulae pertaining to quantum phase fluctuations in 2D; (iv) Discussion of BKT effects in the 2.2 nm film.

**Figure Captions**

**Figure 1|** (a) $R_\square$ vs T for a-MoGe films with different thicknesses. (b) $R_\square^{9K}$ (black curve) and $T_c$ (blue curve) as a function of film thickness ($t$); the solid lines are guide to the eye. *Inset* of (b) shows variation of $R_\square^{9K}$ (black dot) and $\rho$ as a function of $1/t$.

**Figure 2|** (a), (b) *G(V)-V* tunneling spectra at different temperatures for 20 nm and 2 nm thick films respectively. (c), (d), (e) (*left panels*) Intensity plot of $G_N(V)$ as function of bias voltage and temperature for three films with thickness 11 nm, 4.5 nm, 2 nm respectively; temperature variations of $R_\square$ are shown in the same panels; (*right panels*) corresponding spatial maps of $G_N(0)$ maps at 450 mK. (f) Variation of $T_c$ and $T^*$ with film thickness. (g) $\Delta(0)$ and $\Delta(0)/k_BT_c$ as function of $T_c$.

**Figure 3|** (a) $\lambda^{-2}$ as a function of temperature for films with different thicknesses. For 2.2 nm, $\lambda^{-2}$ is multiplied by 2.5 for clarity. Solid lines represent the temperature variation expected from the dirty-limit BCS theory. (b) $\lambda_{expt}^{-2}(T \rightarrow 0)/\lambda_{BCS}^{-2}(0)$ for different thicknesses ($t$). (c) Temperature variation of $\lambda^{-2}$ for 2.2 nm thick sample; the solid straight line is a fit to the linear $T$ region; *inset* expanded view of the linear $T$ variation. (d) $J_s$ and $\Delta$ as a function of thickness.

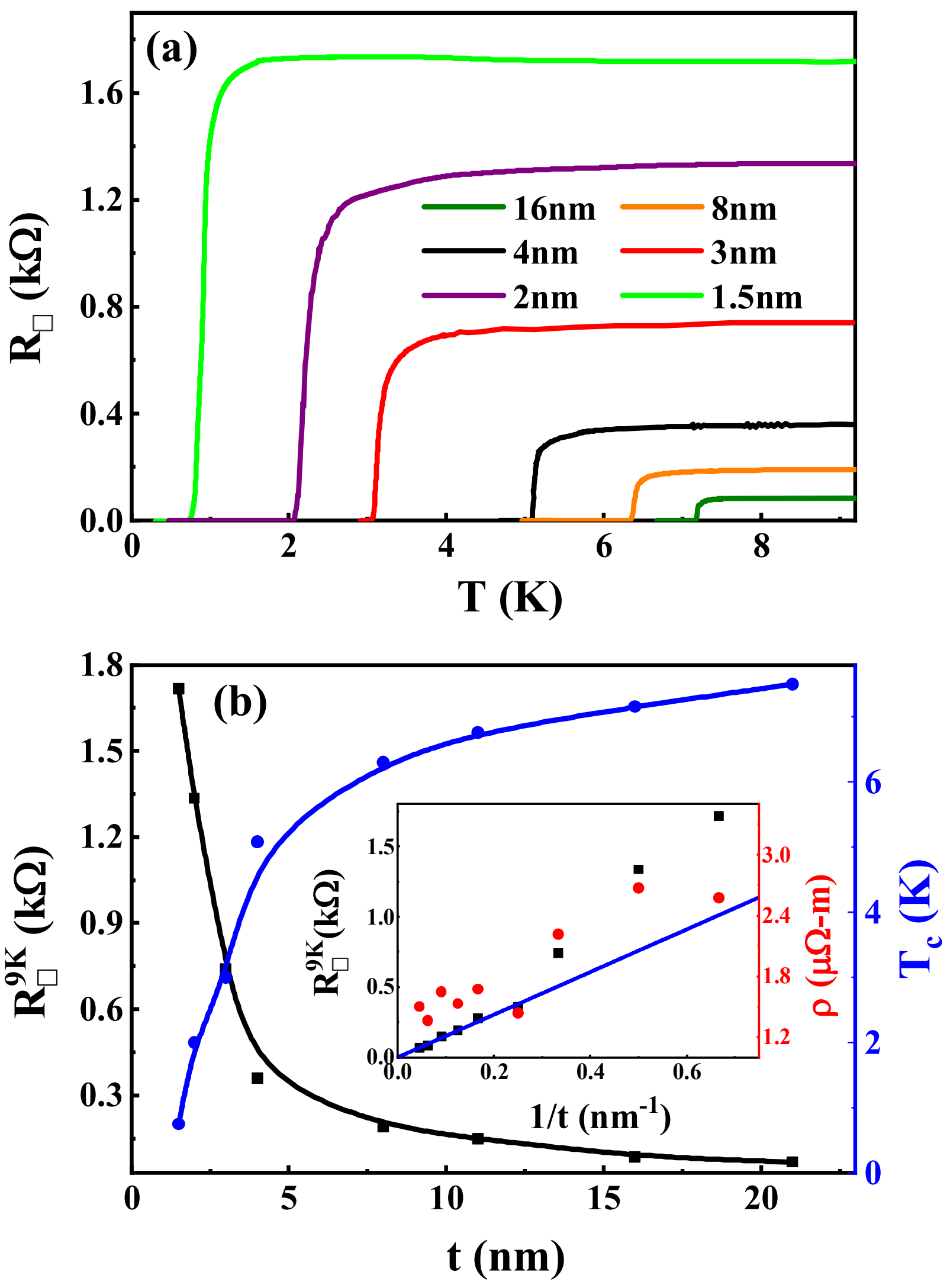

Figure 1

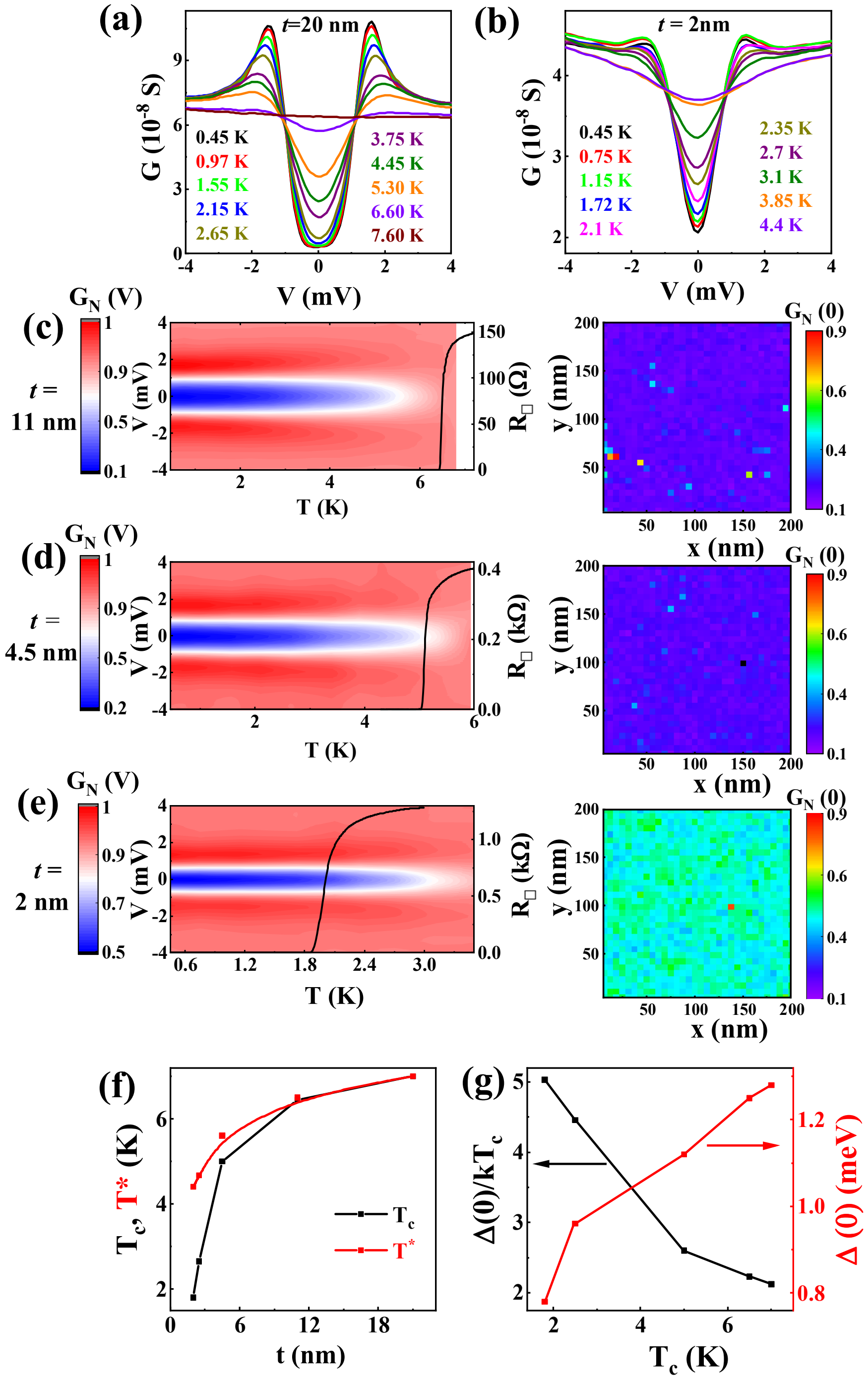

Figure 2

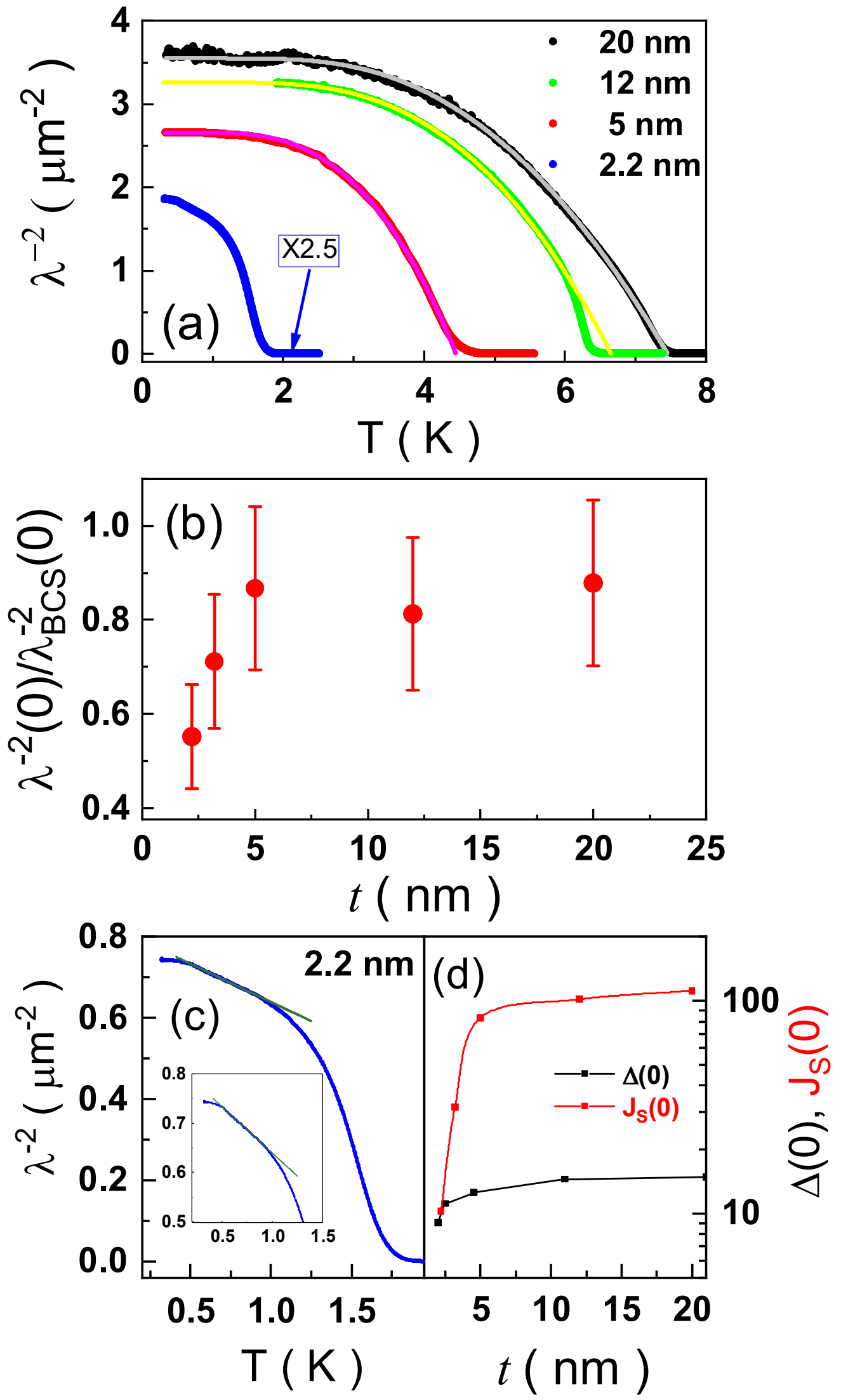

(a)
20 nm
12 nm
5 nm
2.2 nm
X2.5
λ⁻² ( μm⁻² )
T ( K )
(b)
λ⁻²(0)/λ⁻²BCS(0)
t ( nm )
(c)
2.2 nm
λ⁻² ( μm⁻² )
T ( K )
(d)
Δ(0)
JS(0)
Δ(0), JS(0)
t ( nm )

Figure 3

## Supplementary Material

### I. Fitting of the tunnelling spectra and temperature variation of superconducting energy gap

The tunneling conductance between a normal-metal tip and a superconductor is theoretically described by the equation[1],

$$G(V) \propto \frac{1}{R_N}\int_{-\infty}^{\infty} N_s(E)\frac{\partial f(E-eV)}{\partial E}dE, \qquad (1)$$

where $N_s(E) = Re\left(\frac{|E+i\Gamma|}{\sqrt{(|E+i\Gamma|)^2-\Delta^2}}\right)$ is the single particle density of states in the superconductor. When $\Gamma = 0$, this expression reduces to the usual BCS expression. $\Gamma$ is a phenomenological parameter that is incorporated in the density of states to account for possible broadening in the presence of disorder[2]. Fig. S1 shows the fit of the normalized $G_N(V)$-$V$ spectra ( left panels ) along with the temperature variation of $\Delta$ and $\Gamma$ ( right panels ) for the corresponding samples. The temperature variations of $R_\square$ are also shown in the right panels.

### II. Penetration depth measurements from low-frequency mutual inductance technique

The penetration depth of our samples is measured using a low-frequency penetration depth technique[3] operating at 30 kHz. In this measurement a circular superconducting film (diameter 8 mm) is sandwiched between a quadrupolar primary and a dipolar secondary coil (Fig. S2 (a)) and real and imaginary part of the mutual inductance (*M' and M"*) are measured using a lock-in amplifier. In general the magnetic shielding response of the films is described by a complex quantity ($\tilde{\lambda}$) of the form $\tilde{\lambda}^{-2} = \lambda^{-2} + i\delta^{-2}$, where $\lambda$ is the London penetration depth and $\delta$ is the skin depth. To determine $\tilde{\lambda}$ from *M' and M"* we use the following method. We create a lookup table by calculating $M$ for a range of $\lambda^{-2}$ and $\delta^{-2}$, by numerically solving the Maxwell and London equations using finite element analysis. The actual values of $\lambda^{-2}$ and $\delta^{-2}$ are then obtained by comparing the measured values of *M' and M"* with the calculated ones. Further details of this technique is given in refs. 4,5. Fig. S2 (b) shows *M' and M"* measured for different samples ( left panels ) and the corresponding temperature variation of $\lambda^{-2}$ and $\delta^{-2}$ ( right panels ) for different samples.

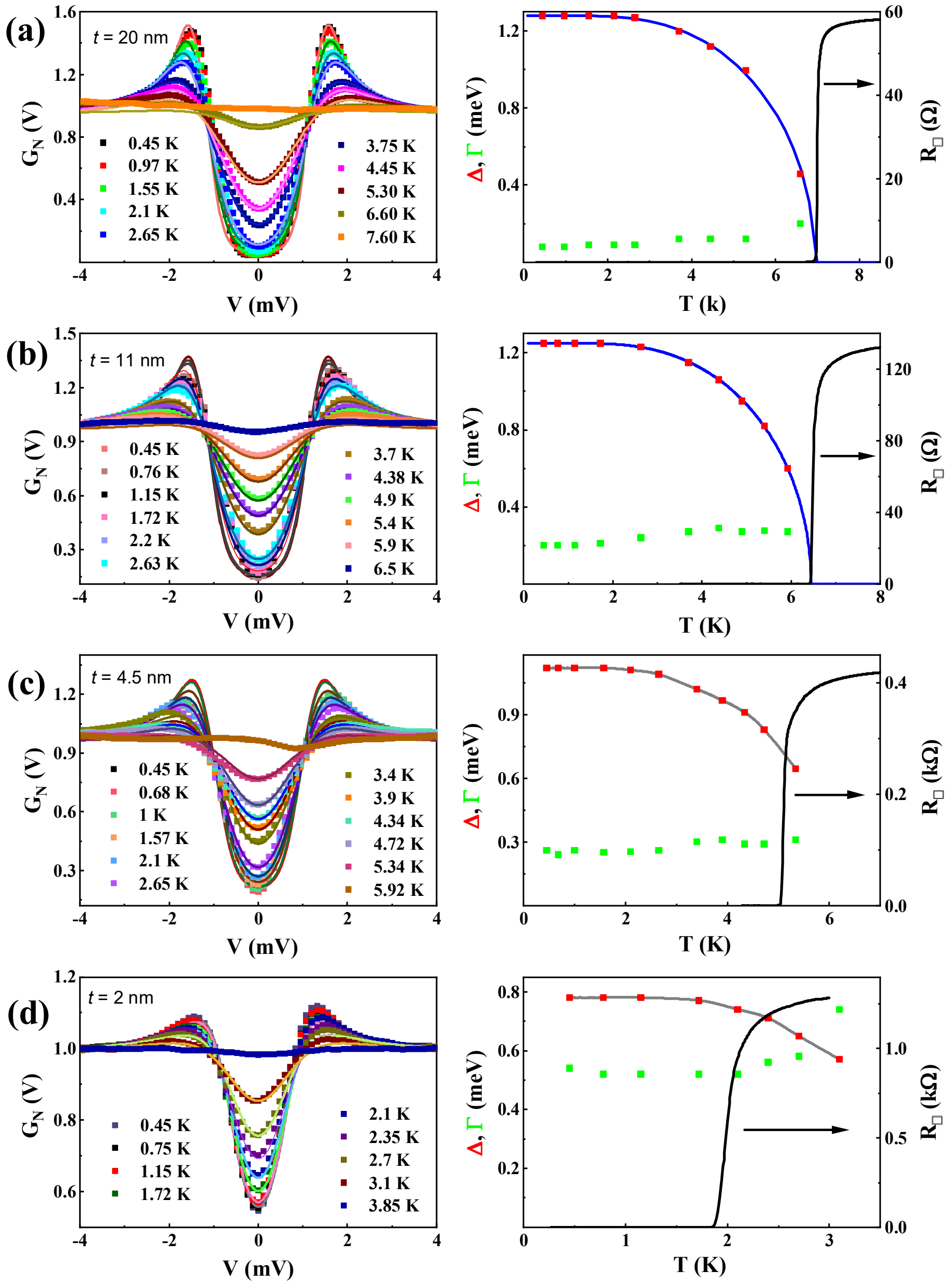

**Fig. S1|** (a)-(d) (left panels) The normalized $G_N(V)$-$V$ spectra along with the fits with the BCS+Γ model for 4 samples with different thickness. (right panels) Temperature variation of the corresponding Δ and Γ extracted from the fits; the temperature variation of the sheet resistance are shown in the same panels. The blue lines in (a) and (b) show the BCS fit to the temperature variation of Δ. The solid grey lines in (c) and (d) are guide to the eye; for these two thickness a pseudogap is observed above $T_c$.

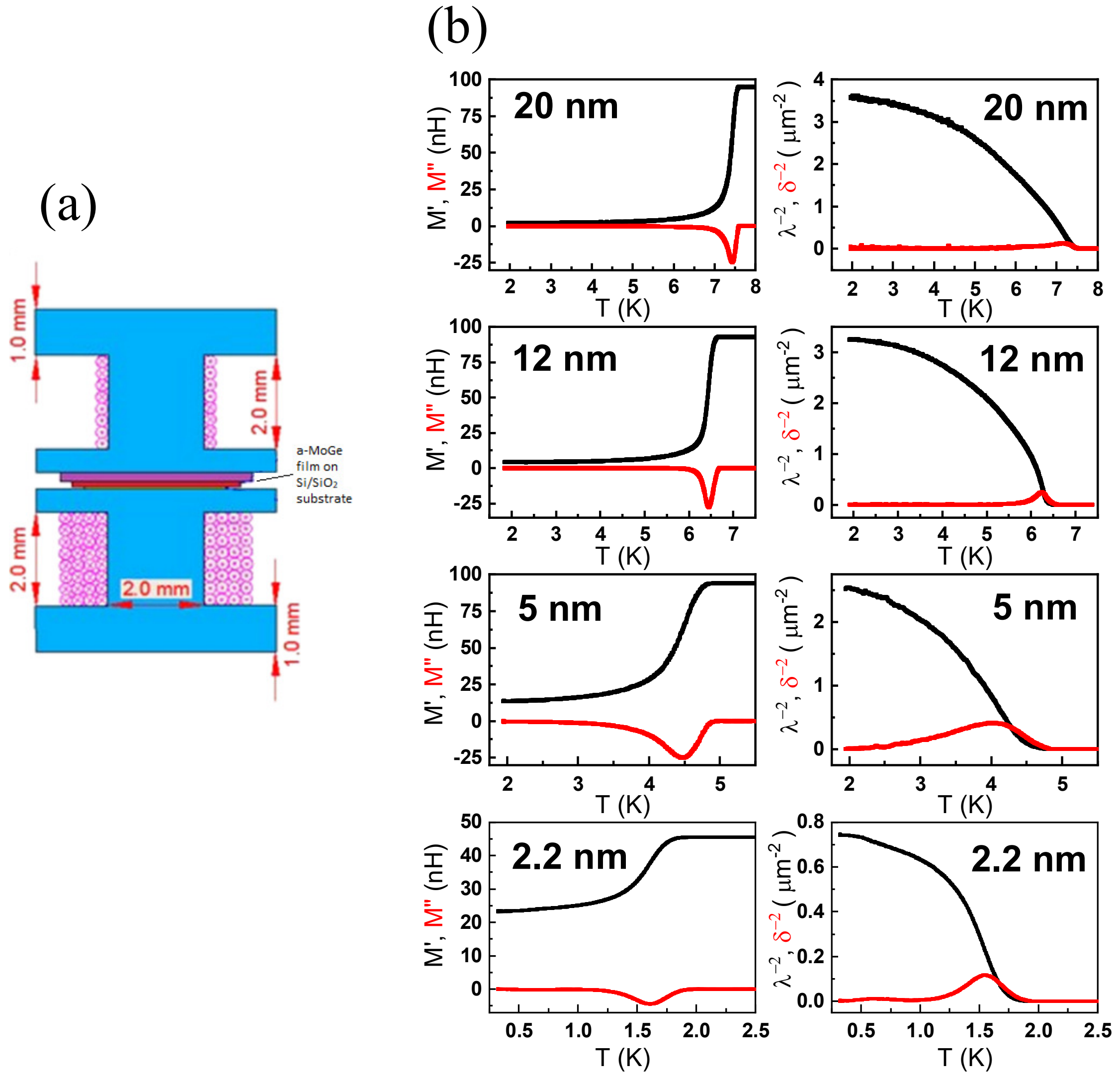

**Fig. S2|** (a) Schematic diagram of the two-coil mutual inductance setup. The superconducting film is sandwiched between a quadrupolar primary coil (top) and a dipolar secondary coil (bottom). (b) (left panels) *M'* and *M"* as a function of temperature for films of different thickness. (right panels) corresponding $\lambda^{-2}$ and $\delta^{2}$ as a function of temperature.

## III. Quantum phase fluctuations in two dimensions

To estimate the role of quantum phase fluctuations we start from the general structure of the quantum phase-only action in *D* dimension, which has been derived e.g. in ref. 6,7,8,9,

$$S_{eff} = \frac{1}{2}\sum_q \left[\frac{\hbar^2\omega_n^2}{4}\tilde{\chi}_{\rho\rho}(q) + \frac{D_s}{4}\boldsymbol{k^2}\right] |\theta(\boldsymbol{q})|^2 \ , \quad (1)$$

Here $q = (i\omega_n, \boldsymbol{k})$, where $i\omega_n = \frac{2\pi k_B T\eta}{\hbar}$ are Bosonic Matsubara frequencies and $\boldsymbol{k}$ is the momentum, $D_s = \frac{\hbar^2 n_s}{m}$, where $n_s$ is the superfluid density (in *D* dimensions), *m* is the effective electron mass, and $\tilde{\chi}_{\rho\rho}$ is the density-density susceptibility dressed at RPA level by the Coulomb interaction *V(**k**)*:

$$\tilde{\chi}_{\rho\rho} = \frac{\chi^0_{\rho\rho}}{1+V(\boldsymbol{k})\chi^0_{\rho\rho}} \ , \quad (2)$$

In Eq. (2) $\chi^0_{\rho\rho}$ represents the bare charge susceptibility, which reduces in the static limit to the compressibility of the electron gas, $\chi^0_{\rho\rho}(\omega_n = 0, \boldsymbol{k} \to 0) \equiv \kappa$. The nature of the Goldstone phase mode is dictated by the form of the charge susceptibility. For the neutral system Coulomb interactions are absent and $\tilde{\chi}_{\rho\rho}$ in Eq. (1) can be replaced by the bare one $\chi^0_{\rho\rho}$. Thus, in the long-wavelength limit the pole of the phase propagator defined by the Gaussian action (1) defines the sound-like dispersion of the Goldstone mode:

$$\omega^2 = \frac{n_s}{m\kappa}\boldsymbol{k}^2 \ , \quad (3)$$

where we performed the analytical continuation $i\omega_n \to \omega + i\delta$ to real frequencies. On the other hand, in the presence of Coulomb interaction the long-wavelength limit of the charge compressibility (2) scales as $\tilde{\chi}_{\rho\rho} \to \frac{1}{V(\boldsymbol{k})}$, so the Gaussian action reads:

$$S_{eff} = \frac{1}{2}\sum_q \frac{1}{4V(\boldsymbol{k})}\left[\hbar^2\omega_n^2 + D_s V(\boldsymbol{k})\boldsymbol{k^2}\right]|\theta(\boldsymbol{q})|^2 = \frac{1}{2}\sum_q \frac{1}{4V(\boldsymbol{k})}\left[\hbar^2\omega_n^2 + \hbar^2\omega_P^2(\boldsymbol{k})\right]|\theta(\boldsymbol{q})|^2, \quad (4)$$

where we defined:

$$\hbar^2\omega_P^2(\boldsymbol{k}) = D_s V(\boldsymbol{k})\boldsymbol{k^2} \ , \quad (5)$$

In the usual isotropic 3D case $V(\boldsymbol{k}) = \frac{e^2}{\varepsilon_0 \varepsilon_B k^2}$, where $\varepsilon_0$ is the vacuum permittivity and $\varepsilon_B$ the background dielectric constant. In this case $\omega_P^2 = \frac{e^2 n_s^{3D}}{\varepsilon_0 \varepsilon_B m}$, which is the usual 3D plasma frequency. Notice that for clean superconductors, $n_s \approx n$ at $T = 0$, so one recovers the same expression as the plasma frequency of the normal state. This has been used in the main text to estimate $\varepsilon_B$ in thick *a*-MoGe films. On the other hand, as discussed in the main text, for a film thickness $t$ smaller than the coherence length we can expect to be in the 2D limit. In this case the Coulomb potential in momentum space reads[8] $V(\boldsymbol{k}) = \frac{e^2}{2\varepsilon_0 \varepsilon_B |\boldsymbol{k}|}$, and one recovers from Eq. (4)-(5) the typical 2D plasma mode:

$$\omega_P^2 = \frac{n_s^{2D} e^2}{2\varepsilon_0 \varepsilon_B m} |\boldsymbol{k}|. \quad (6)$$

Once the spectrum of the phase mode is established, one can estimate the effect of quantum phase fluctuations on the superfluid density due to anharmonic phase interactions beyond the Gaussian level. A simple example of such anharmonic effects comes from the analogy between the superconducting (SC) phase-only model and the classical *XY* model:

$$H_{XY} = -\sum_{ij} J \cos(\theta_i - \theta_j), \quad (7)$$

where $\theta_i$, which is the angular variable for the 2D (planar) spins arranged e.g. on a square lattice, represents the SC phase in the mapping to a SC problem. The *XY* Hamiltonian (7) can be thought as an effective coarse-grained model for the phase degrees of freedom, where $J$ is the effective Josephson coupling between neighboring grains. In the continuum limit in 2D one can approximate the cosine term of Eq. (7) in a power expansion in the phase gradient:

$$H_{XY} \cong -\frac{J}{2} \int d^2x \left[ (\nabla\theta)^2 - \frac{\xi_0^2}{12} \sum_{\alpha=x,y} \left( \frac{\partial\theta}{\partial\alpha} \right)^4 \right]. \quad (8)$$

Here we assumed that the coarse-graining scale coincides with the SC coherence length $\xi_0$, since above it amplitude fluctuations also set in. As one can see, the Gaussian term of Eq. (8) coincides with the gradient term of Eq. (1), allowing one to identify the coupling $J$ with the 2D superfluid stiffness, $J = \frac{\hbar^2 n_s^{2D}}{4m}$. When the anharmonic terms of Eq. (8) are included one can approximate, within perturbation theory with respect to the harmonic term:

$$\frac{\xi_0^2}{12}\sum_{\alpha=x,y}\left(\frac{\partial\theta}{\partial\alpha}\right)^4 \cong \frac{\xi_0^2}{2}\sum_\alpha\left[\left\langle\left(\frac{\partial\theta}{\partial\alpha}\right)^2\right\rangle\left(\frac{\partial\theta}{\partial\alpha}\right)^2\right] = \frac{\xi_0^2}{2}\langle(\nabla\theta)^2\rangle(\nabla\theta)^2 \qquad (9)$$

As a consequence, by denoting with $n_s^0$ the bare 2D stiffness at harmonic level, the renormalized stiffness $n_s$ which takes into account the anharmonic corrections reads:

$$\frac{n_s}{n_s^0} = 1 - \frac{\xi_0^2\langle(\nabla\theta)^2\rangle}{4} = 1 - \frac{\langle(\Delta\theta)^2\rangle}{4}\ , \qquad (10)$$

where, $\langle(\Delta\theta)^2\rangle = \xi_0^2\langle(\nabla\theta)^2\rangle$. On more general ground, if one accounts for the full anharmonic structure of the cosine term in Eq. (7) the superfluid density with the self-consistent harmonic approximation reads[8]:

$$\frac{n_s}{n_s^0} = Exp\left[-\frac{\langle(\Delta\theta)^2\rangle}{4}\right]. \qquad (11)$$

Within the classical *XY* model (7) the average of the phase gradient computed at Gaussian level simply scales as $\langle(\Delta\theta)^2\rangle = k_B T/J$, so that Eq. (10) accounts for the linear depletion of the superfluid density at low temperature reported in Monte Carlo simulations of the classical *XY* model[10,11].

The simplest approach to account for anharmonicity within a quantum phase-only model is to map the SC problem into a quantum *XY* model[6,7,8,9]. This is equivalent to retaining the structure (11) for the anharmonic corrections to the superfluid density and computing the average $\langle(\nabla\theta)^2\rangle$ with the quantum phase-only action (1). This has two main consequences: (i) the quantum model allows for finite corrections also at $T = 0$; (ii) the classical limit $\langle(\Delta\theta)^2\rangle = k_B T/J$ is only attained above a certain crossover temperature. Then, by using the model (4) one easily obtains:

$$\langle(\Delta\theta)^2\rangle = \xi_0^2\langle(\nabla\theta)^2\rangle = \frac{\xi_0^2 T}{N}\sum_{i\omega_n,\boldsymbol{k}} 4\boldsymbol{k}^2 V(\boldsymbol{k})\frac{1}{\omega_n^2+\omega_p^2(\boldsymbol{k})} = \frac{\xi_0^2}{V}\sum_{\boldsymbol{k}} 4\boldsymbol{k}^2 V(\boldsymbol{k})\frac{1}{2\omega_p(\boldsymbol{k})}\left[1 + 2b\big(\omega_p(\boldsymbol{k})/T\big)\right] \quad (12)$$

where $N$ is the number of lattice sites and $b(x)$ is the Bose function. At $T \rightarrow 0$ the Bose factor vanishes and one can easily compute the quantum correction for the 2D plasmon as:

$$\langle(\Delta\theta)^2\rangle(T = 0) = \frac{\xi_0^2}{V}\sum_{i\omega_n,\boldsymbol{k}} 4\boldsymbol{k}^2 V(\boldsymbol{k})\frac{1}{2\omega_p(\boldsymbol{k})} = \xi_0^2\sqrt{\frac{2e^2}{\varepsilon_0\varepsilon_B D_s}}\int\frac{d^2\boldsymbol{k}}{(2\pi)^2}\sqrt{|\boldsymbol{k}|} = \xi_0^2\sqrt{\frac{2e^2}{\varepsilon_0\varepsilon_B D_s}}\frac{k_0^{5/2}}{5\pi} \quad (13)$$

The upper cut-off $k_0$ is connected to the coarse graining scale $\xi_0$. We can roughly take $k_0 \sim 1/\xi_0$ up to a multiplicative factor of the order of unity. We then get,

$$\langle(\Delta\theta)^2\rangle(T=0) = \frac{1}{5\pi}\sqrt{\frac{E_c}{J}} , \quad (14)$$

where we introduce the Coulomb energy scale given by,

$$E_c = \frac{e^2}{2\varepsilon_0\varepsilon_B\xi_0} . \qquad (15)$$

As we mentioned above, the superfluid stiffness $J$ is connected to the effective 2D superfluid density by, $J = \frac{D_s}{4} = \frac{\hbar^2 n_s^{2D}}{4m}$. In the case of thin films of thickness $t$ it is then connected to the measured penetration depth $\lambda$ as:

$$J = \frac{\hbar^2 n_s^{2D}}{4m} = \frac{\hbar^2 t}{4\mu_0 e^2\lambda^2} \quad . \quad (16)$$

By using the estimates of $E_c$ and $J$ reported in the main text we obtain $\frac{n_s}{n_s^0} \approx 0.78$. While this is somehow larger than the experimental value, we can nonetheless argue that it is of the correct order of magnitude. It is worth noting that by deriving microscopically anharmonic terms in the phase degrees of freedom one can also get quantum corrections not included in the quantum *XY*-model approximation[9]. In general, the quantum corrections within the *XY* model are larger than the ones obtained within the microscopically derived anharmonic phase-only action, putting the present estimate in the correct ballpark.

Finally, we should also consider that disorder and dissipation can also affect the above estimate (14) in different ways. From one side, disorder can trigger inhomogeneity on the local SC properties, as shown in Fig. 2e of the manuscript where we report the spatial variation of the tunneling conductance for the thinnest film. Within an effective *XY* model approach, this can be modelled as an inhomogeneity of the local SC stiffness, i.e. of the local SC couplings of the model (7):

$$H_{XY} = -\sum_{ij} J_{ij}\cos(\theta_i - \theta_j). \quad (17)$$

In this situation the superfluid density at $T = 0$ is lower than the average value $J = \langle J_{ij}\rangle$ by a quantity that scales for uncorrelated disorder with the variance of the SC couplings[10,11]. This additional suppression of the superfluid density due to inhomogeneity reflects in the emergence of a finite-frequency absorption that is expected to occur at relatively low-energies, i.e. below the $2\Delta$ threshold for quasiparticle absorption in dirty superconductors[12,13]. This effect has been indeed observed in strongly disordered $InO_x$ and NbN films[14,15]. Such a low-frequency extra absorption can also have a feedback on the quantum phase mode. Indeed, as discussed in Ref. 8 within the context of $d$-wave superconductors, where such effect is present even in the absence of disorder, by including dissipative effects the estimate (14) of quantum corrections for the homogeneous model gets in general reduced, and the crossover to the classical regimes occurs at progressively smaller temperature scales. Both effects could then be in principle relevant for our thinnest $a$-MoGe films, but a quantitative estimate would require the precise knowledge of the stiffness inhomogeneity and of the extra absorption, that is not available so far.

**IV. Berezinskii-Kosterlitz-Thouless transition in 2.2 nm thick film**

As we mentioned in the main text, even for our thinnest film ( ~ 2.2 nm ) the low-temperature stiffness $J(T \to 0) \approx 10$ K given by Eq. (16) is still large enough that the Berezinskii-Kosterlitz-Thouless ( BKT ) temperature $T_{BKT}$ is very close to the BCS mean-field temperature ( $T_{BCS}$ ). In addition, the sample inhomogeneity emerging at strong disorder is expected to smear out the BKT signatures, as for example the universal jump of the superfluid stiffness[10], as already observed experimentally in thin films[16,17] of NbN. To quantify these effects we analyzed the temperature dependence of the superfluid stiffness in the thinnest film within the same scheme described in Ref. 16,17.

As discussed in the main text, both quasiparticle excitations and phase fluctuations contribute to the depletion of $J$ towards zero. The former effect can be captured by a BCS-like fit which works pretty well at intermediate temperatures, except for the low-$T$ region where the anomalous linear behavior has been reported. The superfluid stiffness can then be fitted by the dirty-limit expression,

$$\frac{J^{BCS}(T)}{J^{BCS}(0)} = \frac{\Delta(T)}{\Delta(0)} \tanh\left[\frac{\Delta(T)}{2k_B T}\right], \qquad (18)$$

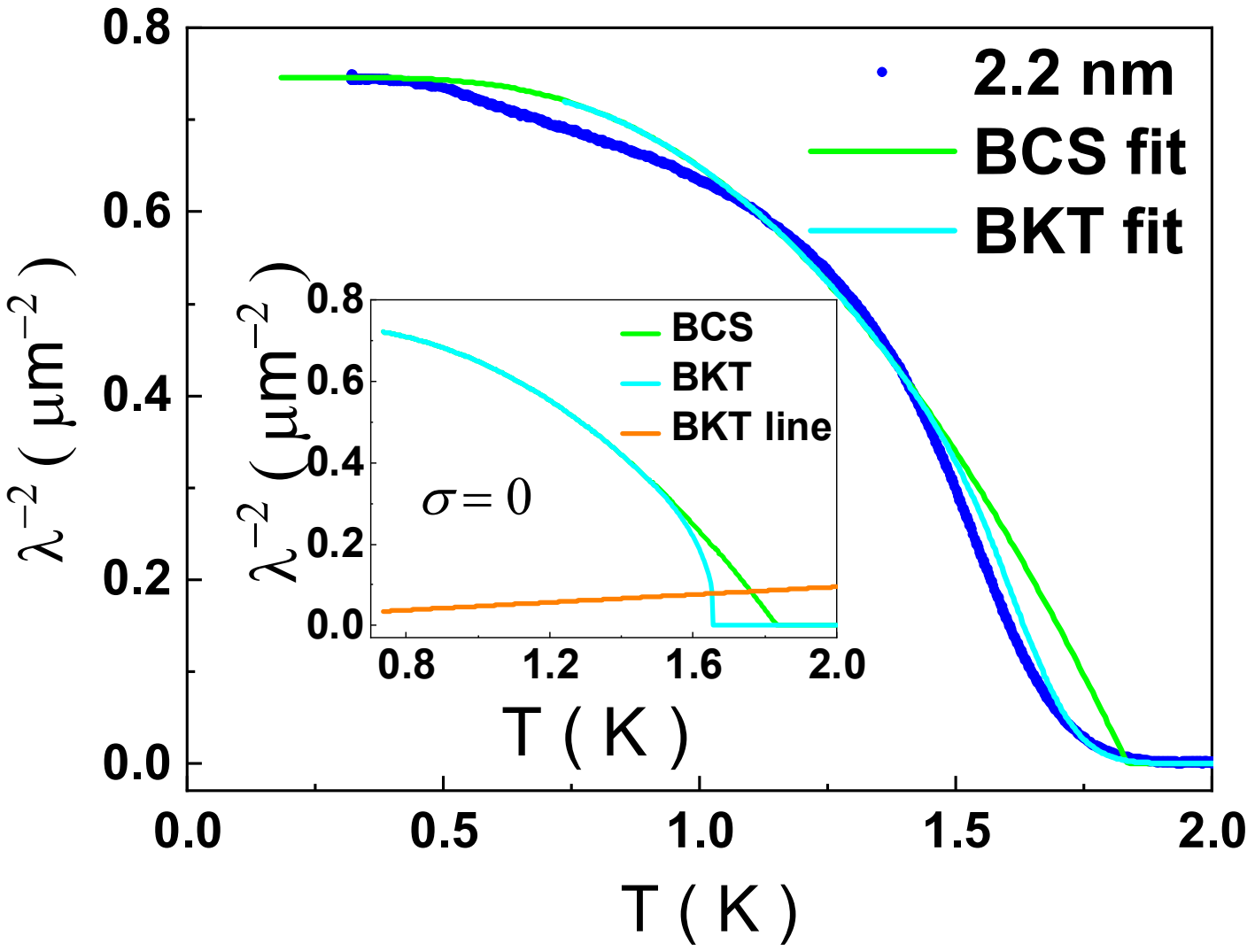

**Fig. S3|** Temperature variation of $\lambda^{-2}$ for the 2.2 nm thick sample along with the fits from dirty-limit BCS formula and BKT theory; The *inset* shows the corresponding BCS and BKT variation for the hypothetical case of zero disorder, i.e. $\sigma = 0$ . The universal BKT line is shown in orange.

by using eventually $\frac{\Delta(0)}{T_{BCS}}$ as a free parameter, to account for the relatively large $\frac{\Delta(0)}{k_B T_c}$~2 ratio extracted by the STM analysis. Notice that here $\tilde{J}^{BCS}(0)$ already includes the corrections due to longitudinal phase fluctuations at $T = 0$ discussed in the main text.

For what concerns transverse ( i.e. vertical ) phase fluctuations their effect will be accounted for by numerical solution of the BKT renormalization-group (RG) equations, whose relevant variables are the dimensionless quantities:

$$K(0) = \frac{\pi J^{BCS}(T)}{k_B T}, \qquad (19)$$

$$g(0) = 2\pi e^{-\mu/k_B T}, \qquad (20)$$

where $\mu$ is the free energy of a vortex core, with radius equal to the coherence length $\xi$, and $g$ is the vortex fugacity. Notice that $J^{BCS}(T)$ enters here to determine the initial value of $K$, i.e. its short-distance value. Its long-distance value follows from the solution of the well-known RG equations[18,19,20,21,22]:

$$\frac{dK}{dl} = -K^2 g^2, \qquad (21)$$

$$\frac{dg}{dl} = (2 - K)g, \qquad (22)$$

where $l \equiv ln(r/\xi_0)$ is the rescaled length scale. The observed superfluid density is identified by the limiting value of $K$ as one goes to large distances,

$$J \equiv \frac{k_B T K(l\to\infty)}{\pi}. \qquad (23)$$

From Eqs. (21)-(22) one sees that when $g \to 0$ single-vortex excitations are ruled out from the system, which is then SC: indeed, when $g \to 0$, $K$ goes to a constant and then $J$ from Eq.(23) is finite. If instead $g \to \infty$ at large distances it means that vortices proliferate and drive the transition to the non-SC state, since $K \to 0$. The large-scale behavior depends on the initial values of the coupling constants $K, g$, which in turn depend on the temperature. The BKT transition temperature is defined as the highest value of $T$ such that $K$ flows to a finite value, so that $J$ is finite. This occurs at the fixed point $K = 2, g = 0$, so that at the transition one always has:

$$K(l \to \infty, T_{BKT}) = 2 \;\Rightarrow\; \frac{\pi J(T_{BKT})}{k_B T_{BKT}} = 2 \;, \qquad (24)$$

while above it, $J = 0$. As a consequence, in the ideal system $J$ jumps discontinuously at $T_{BKT}$ from the universal value $\frac{2k_B T_{BKT}}{\pi}$ to zero. In addition, already before $T_{BKT}$ bound vortex-antivortex pair contribute to deplete $J$ with respect to its BCS estimate (18). This effect is usually negligible when $\mu$ is large, as it is the case within the standard *XY* model[8,22], where $\mu_{XY} \sim \left(\frac{\pi^2}{2}\right) J$. However, in thin films of ordinary superconductors $\mu$ is usually quite smaller, leading a significant renormalization of $J$ due to bound vortex-antivortex pairs already before $T_{BKT}$, as observed e.g. in NbN films[16,17]. Finally, to account for the sample inhomogeneity observed by STM we will follow the same procedure suggested in Ref. 16,17. We will assume that the local BCS stiffness $J_i^{BCS}$ is distributed according to a given probability density $P(J_i)$, and that the local BCS transition temperature $T_{BCS}^i$ scale accordingly. Then by solving numerically the RG equations (21)-(22) we can compute the local stiffness $J_i(T)$. The overall superfluid stiffness is then computed phenomenologically as an average value $J_{av}$:

$$J_{av}(T) = \sum_i P(J_i) J_i(T) \qquad (25)$$

For the sake of concreteness we will use for $P(J_i)$ a Gaussian distribution centered around $J_0$:

$$P(J_i) = \frac{1}{\sqrt{2\pi}\sigma} exp\left[-\frac{(J_i - J_0)^2}{2\sigma^2}\right]. \qquad (26)$$

When all the stiffness $J_i(T)$ are different from zero, as is the case at low temperatures, the average stiffness will be centered around the center of the Gaussian distribution (26), so that it will coincide with $J_0(T)$. However, by approaching $T_{BKT}$ defined by the average $J_0(T)$ not all the patches make the transition at the same temperature, so that the BKT jump is rounded and $J_{av}$ remains finite above the average $T_{BKT}$. This leads to a rather symmetric smearing of the superfluid-density jump with respect to the abrupt downturn observed for the clean case, that is progressively more pronounced for increasing $\frac{\sigma}{J_0\,(T=0)}$. It is worth noting that such a phenomenological approach accounts rather well for experiments in thin films on NbN [16,17], and it has been recently validated theoretically by Monte Carlo simulations within a inhomogeneous 2D *XY* model in the presence of correlated disorder[10].

In Fig. S3 we show the result of the above fitting procedure for the 2.2 nm thick $a$-MoGe film. The curve labeled BCS represents the BCS fit of the average stiffness based on Eq. (18) above. From the fit we obtain the values $T_{BCS} = 1.84$ K and $\frac{\Delta(0)}{k_B T_{BCS}} = 1.9$, that is consistent with the STM estimate. As mentioned above, the BCS fit cannot capture the linear depletion at low temperatures, but it captures rather well the overall suppression due to quasiparticle excitations up to a temperature $T \sim 1.5$ K. Here the BKT curve starts to deviate from the BCS one due to the effect of bound vortex-antivortes pairs, as it is evident in the inset where we show the homogeneous case ($\sigma = 0$). Here we used $\mu/J_0 = 1.3$, as estimated from the comparison with the experimental data in the main panel. Such a low vortex-core energy value leads to a significant renormalization of the stiffness already below the universal jump, which always occurs, according to Eq. (24), at the intersection with the universal $\frac{2k_BT}{\pi}$ line, i.e. at $T_{BKT} = 1.65$ K in our case. The abrupt jump is however smeared out in the presence of inhomogeneity, as shown in the main panel, where the BKT line corresponds to the average procedure encoded in Eq. (25) for $\sigma/J_0 = 0.05$. The overall fit reproduces reasonably well the experimental trend, with a global transition temperature $T_c \cong 1.8$ K. Overall, we can conclude that the BKT analysis further supports the conclusion that our thinnest sample is in the 2D limit. However, the manifestation of BKT effects is partly blurred out by the inhomogeneity, which unavoidably comes along with the increase of effective disorder in the thinnest samples.